# Plasmonic Vortices Host Magnetoelectric Interactions

Atreyie Ghosh,[1,2] Sena Yang,[1] Yanan Dai,[1,3,4*] W. Vincent Liu,[1] and Hrvoje Petek[1*]

[1]Department of Physics and Astronomy, University of Pittsburgh and IQ Initiative, Pittsburgh, PA 15260, USA

[2]James Franck Institute, University of Chicago, Chicago, IL 60637, USA

[3]Department of Physics, Southern University of Science and Technology (SUSTech), Shenzhen, 518055, China

[4]Quantum Science Center of Guangdong-Hong Kong-Macao Greater Bay Area (Guangdong), Shenzhen, 518045, China

**ABSTRACT**

The vector $E\times H$ and pseudoscalar $E\cdot H$ products of electric and magnetic fields are separately finite in vacuum transverse electric and magnetic (TEM) plane waves, and angular momentum structured light. Current theories of interactions beyond the standard model of particle physics invoke $E\cdot H\neq 0$ as the source term in the axion law that describes interactions with the cosmological dark matter axion particles outside of the quartet of Maxwell's equations. $E\cdot H\neq 0$ also drives relativistic spin-charge magnetoelectric excitations of axion quasiparticles at a distinctively higher condensed matter scale in magnetic and topological materials. Yet, how to drive coherent $E\cdot H$ responses is unknown, and provides motivation to examine the field polarizations in structured light on a deep sub-diffraction limited spatial scale and sub-optical cycle temporal scale by ultrafast nonlinear photoemission electron microscopy. By analytical theory and ultrafast coherent photoemission electron microscopy, we image $E\cdot H$ fields in surface plasmon polariton vortex cores at subwavelength scales, where we find that the magnetoelectric relative to the dipole density is intensified on a ~10 nm diameter scale as a universal property of plasmonic vortex fields. The generation and nanoscale localization of $E\cdot H$ fields introduces the magnetoelectric symmetry class, having the parity $\mathcal{P}$ and time reversal $\mathcal{T}$ broken, but the joint $\mathcal{PT}$ symmetry preserved. The ability to image the optical fields of plasmonic vortex cores opens the research of ultrafast microscopy of magnetoelectric responses and interactions with axion quasiparticles in solid state materials.

*e-mail: daiyn@sustech.edu.cn; petek@pitt.edu

## I. INTRODUCTION

Defining the electric ($E$) and magnetic ($H$) interactions of electron charge and spin, the four Maxwell's equations have stood the test into the quantum age. The magnetoelectric (ME) effect, originally predicted by Curie[1] and Debye[2], however, generates colinear magnetization and polarization[3] that is beyond the standard Maxwell's equations. Such responses are known to occur in bianisotropic Tellegen materials through the pseudoscalar $\boldsymbol{E}\cdot\boldsymbol{H}\neq\boldsymbol{0}$ term that excites electric charge and spin to break both the parity ($\mathcal{P}$) and time reversal ($\mathcal{T}$), but preserves the joint $\mathcal{PT}$ symmetry. The $\mathcal{PT}$ symmetry has a close correspondence to the elementary particle chiral U(1) symmetry[4], where it explicates the puzzling strong interactions beyond the Standard Model in quantum chromodynamics that is referred to as the "strong charge-parity ($\mathcal{CP}$) problem". Invoking the pseudoscalar fundamental boson axion particle[5,6] provided a resolution to the $\mathcal{CP}$ problem, later[7] becoming a natural candidate for cosmological dark matter[8,9]. The two pseudoscalar terms, the axion field and $\boldsymbol{E}\cdot\boldsymbol{H}$, remarkably arise from seemingly unrelated origins, yet couple naturally by symmetry. Therefore, developing methods for interaction of $\boldsymbol{E}\cdot\boldsymbol{H}$ fields with matter is of fundamental interest, with potential application to electron spin-charge entanglement with potential application in quantum computing, and as interaction in topological physics[10]. Symmetry breaking by $\boldsymbol{E}\cdot\boldsymbol{H}$ fields, referred to as false chirality in the chemical and biological contexts, has also been postulated as a possible source of homochirality of living matter[11]. ME polarizability, $\alpha_{me}$, however, leads to a quantized Hall conductance in transport measurements[12-15], antiferromagnets[16], e.g., $Cr_2O_3$[17], multiferroics[18], strong topological insulators[19] and metamaterials[20]. In application of static fields, however, such ME interactions are incoherent in time and delocalized in space; this provides a strong motivation to develop coherent ultrafast microscopy of the ME effect. Coherent ME response to $\boldsymbol{E}\cdot\boldsymbol{H}$ field has only been reported for $\mathcal{PT}$ symmetry invariant chiral Raman scattering from plasmonic nanoparticles[21].

We report generation and imaging of coherent $\boldsymbol{E}\|\boldsymbol{H}$ fields approaching PHz frequencies for driving and imaging coherent ME interactions and axion physics in solids with nanofemto resolution. We show how $\boldsymbol{E}\|\boldsymbol{H}$ fields arise as a universal property of 2D structured light within phase singularities of plasmonic vortices having orbital angular momentum (OAM) $|l|=1$[22,23].

The axion law describes the pseudoscalar $\boldsymbol{E}\cdot\boldsymbol{H}$ as a source-sink of the axion field $\theta$[24]:

$$(\Box + m_a^2)\theta = -\kappa \boldsymbol{E}\cdot\boldsymbol{B}, \qquad (1)$$

whose space-time oscillation is described by $\Box = \nabla^2 - \frac{1}{c^2}\frac{\partial^2}{\partial t^2}$, where $m_a$ and $\kappa$ are the unknown axion mass and coupling constant, respectively. Axion quasiparticles are excited in topological solids modulus $2\pi$ for $\theta=\pi$[12,25]. In a dynamical context, $\theta$ is a continuous variable[26], that couples nonlinearly to the external ME polarization[26]. The ME effect modifies the Maxwell's equations, however, such as the Gauss's law for magnetism, which becomes $\nabla\cdot(c\boldsymbol{B}+\kappa\theta\boldsymbol{E}) = c\mu_0\rho_m$. In this expression the interaction of axial $\boldsymbol{B}$ with the polar $\boldsymbol{E}$ is mediated by $\theta$ with its strength is defined by the magnetic monopole density $\rho_m$. Therefore, we address how to generate intense, localized, time-dependent coherent $\boldsymbol{E}\|\boldsymbol{H}$ fields with corresponding $\rho_m$ as a source of electron charge-spin entanglement[27], and excitation source of axion physics in the solid state.

We first describe the generation of coherent $\boldsymbol{E}\cdot\boldsymbol{H}$ fields in plasmonic vortex cores. Fig. 1 shows the calculated spatial distribution of $\boldsymbol{E}\cdot\boldsymbol{H}$ field sign for a plasmonic vortex skyrmion field[22], where the $\boldsymbol{E}$ and $\boldsymbol{H}$ vectors rotate within a $\sim\lambda_{\text{SPP}}/2$ diameter ($\lambda_{SPP}=530$ nm is the wavelength of the plasmonic field) area centered on the plasmonic vortex core at the optical frequency, $\omega_L$. This stationary field vector rotation in space contrasts the delocalized amplitude oscillation of propagating TEM waves. At the vortex core, the $\boldsymbol{E}$ and $\boldsymbol{H}$ fields generated by the left circularly polarized (LCP) light are antiparallel, with the relative angle distribution of the $\boldsymbol{E}\cdot\boldsymbol{H}$ field rotating counterclockwise at $\omega_L$, repeating in orientation at $4\omega_L$. Excitation by right circularly polarized light reverses the sign and rotation of the pseudoscalar $\boldsymbol{E}\cdot\boldsymbol{H}$. Going outward, the field orientation abruptly becomes parallel, gradually returning to antiparallel in rings that increase by ~$\lambda_{SPP}/2$ in radius. This cycling between anti-parallel and parallel orientation with a ~$\lambda_{SPP}/2$ period forms a time-averaged target pattern (Fig. 1b). The resulting spin texture of the rotating vortex $\boldsymbol{E}$ and $\boldsymbol{H}$ fields has been identified as that of a magnetic monopole[22].

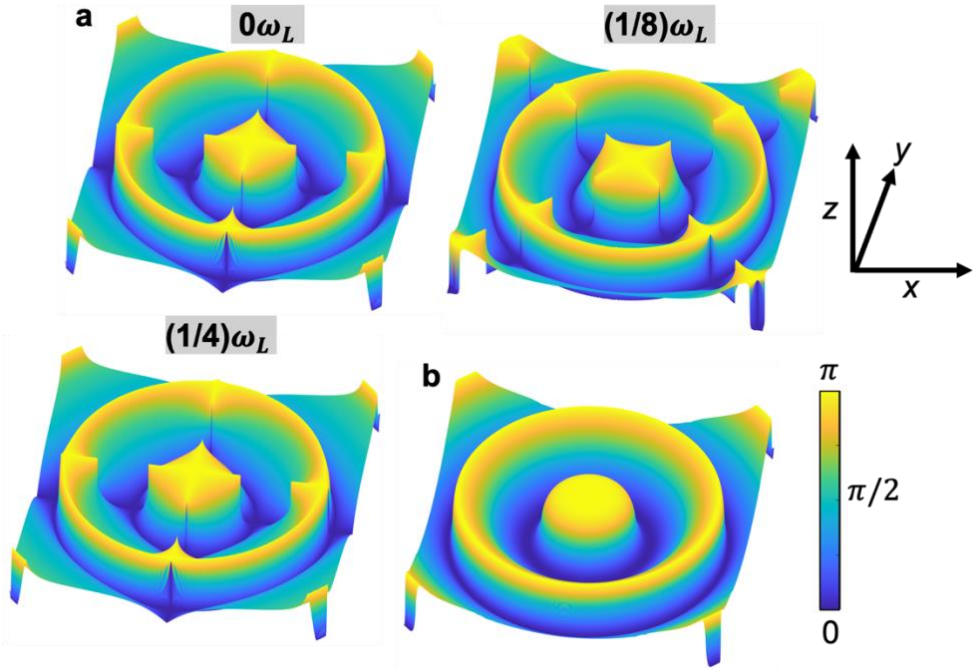

**Fig. 1: The focus region and rotation of $\boldsymbol{E}\cdot\boldsymbol{H}$ fields at the core of an ℓ=-1 surface plasmon vortex. a,** Color maps show how the relative $\boldsymbol{E}\measuredangle\boldsymbol{H}$ angle evolves between 0 and $\pi$ in 1/8 of an optical cycle steps. For LCP generation, in the center $\lambda_{SPP}/2$ diameter region (yellow), the $\boldsymbol{E}$ and $\boldsymbol{H}$ fields remain dominantly antiparallel as they jointly rotate about the vortex core (center) at frequency, $\omega_L$. Progressing outward by $\lambda_{SPP}/2$ the fields become parallel (dark blue) and again antiparallel in a fourfold symmetric structure that repeats at $4\omega_L$. **b,** Averaging the angle over $\omega_L$ illustrates that $\boldsymbol{E}$ and $\boldsymbol{H}$ fields form antiparallel and parallel rings with abrupt realignment for every increase in the radius by $\lambda/4$, outside of the central core. The field structure shows that the ME polarization is generated with singular variations of the sign passing through rings of orthogonal polarization (green). Ultrafast photoelectron microscopy measures the action of such field textures with sub-diffraction limited resolution. The lateral side dimension of each image is $2\lambda_{SPP}$ (1060 nm).

### A. Magnetoelectric density

We will show that within the vortex core the ME interaction dominates over that of the electric ($E^2$) and magnetic ($H^2$) dipoles on a deep subwavelength ~10 nm scale with a universal structure of vortex fields. To evaluate the ME interaction strength relative to that of $E^2$ and $H^2$, we calculate the ME density as defined by Bliokh *et al.*[28]. Motivated by the "superchiral" configuration of Tang and Cohen for studies of molecular chirality[29], they proposed a "supermagnetoelectric"[28] configuration where counterpropagating TEM fields have $E \times H = 0$ and $E \cdot H$ maxima. They defined the ME density as,

$$A_{ME} = \frac{2Re(E^* \cdot H)}{(|E|^2 + \delta |H|^2)}, \qquad (2)$$

where $0 < \delta = \frac{Im(\alpha_m)}{Im(\alpha_e)} \ll 1$ is ratio of the imaginary magnetic $\alpha_m$ to electric $\alpha_e$ polarizabilities. We adopt $A_{ME}$ as a quantifier of ME detectability for recording and imaging the optical ME dressing of solids, which we refer to as Poincaré engineering[23]. Similarly, replacing the numerator in Eq. 2 with $-2Im(E^* \cdot H)$ obtains the chirality density $C_{ME}$[28].

$A_{ME}$ of structured light in form of surface plasmon polariton (SPP) vortices is intrinsically related to its angular momentum, in other words, its topological charge. Light carries spin (SAM) $\sigma$ and OAM $l$, as well as geometric phase[30], degrees of freedom. SAM of evanescent SPP waves is locked transverse to its propagation Poynting vector ($k_{SPP}$)[31]. With total angular momentum being conserved, SPPs propagate and interfere, such that the photonic spin-orbit interaction (SOI) forms spin distributions with quantized topology[32]. Fields are classified by their topological charge, *n*, corresponding to how many times their order parameter wraps a Poincaré sphere. The charge for trivial TEM waves is *n*=0, while the topological meron and skyrmion fields, respectively, have *n*+1/2 and *n*, where *n*≠0 is an integer[23]. In dressing matter such structured fields[33] and polaritons[34] break the $\mathcal{P}$ and $\mathcal{T}$, but preserve $\mathcal{PT}$ symmetry at the quantum scale[10,23,28,35].

## II. QUANTIFYING MAGNETOELECTRIC FIELDS IN PLASMONIC VORTICES

We first examine the $Re(\boldsymbol{E}^* \cdot \boldsymbol{H})$ and the implied $A_{ME}$ as defining properties of plasmonic vortices by simulating them from an analytical model for LCP light ($\sigma = -1$) interacting with coupling structures (CSs) to provide momentum to launch SPP fields with $k_{SPP}$ directed normal to their edges at silver/vacuum interface. The helicity of the excitation light and the geometric charge $m = 0$ of the CSs generates OAM, $l = \sigma + m = -1$. Details of the analytical modeling of plasmonic vortex fields are described in the Appendix A. Fig. 2a shows a calculated color map depicting the spatial variation of $Re(\boldsymbol{E}^* \cdot \boldsymbol{H})$ for illumination of a circular ($Q = \infty$) and square ($Q = 4$) CSs, with $Q$ indicating the number of edges of the polygonal CS. The $Q = \infty$ and 4 CS generate a single vortex and a square vortex array, respectively (details in Fig. A1). The array vortices have the opposite field circulation that alternates between clockwise and counterclockwise from one $\lambda_{SPP}/2$ square domain to the next neighboring square domain[39]. The distribution of $Re(\boldsymbol{E}^* \cdot \boldsymbol{H})$ illustrate that each vortex is a focus of ME interaction, where the parallel (+) or antiparallel (-) alignment of the in-plane electric $\boldsymbol{E}_\parallel$ and magnetic $\boldsymbol{H}_\parallel$ fields (the black and green arrows, respectively) defines its sign.

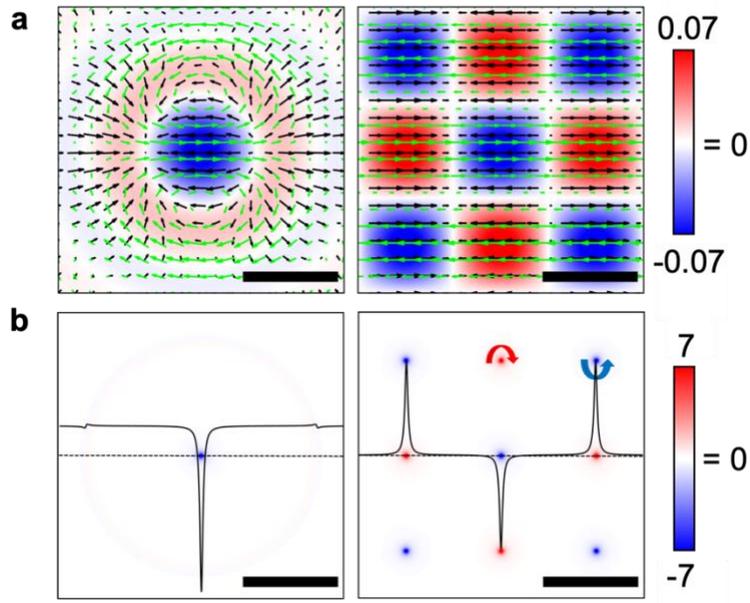

**Fig. 2: The plasmonic fields generating the ME density in SPP vortex cores**. **a**, Color maps of the instantaneous $Re(\mathbf{E}^* \cdot \mathbf{H})$ distribution with black and green arrows denoting the instantaneous SPP $\mathbf{E}_{\parallel}$ and $\mathbf{H}_{\parallel}$ fields, respectively, generated by illuminating circular and square CS with LCP light. The figures show that $Re(\mathbf{E}^* \cdot \mathbf{H})$ has the maximum values at vortex cores where $\mathbf{E}_{\parallel}$ and $\mathbf{H}_{\parallel}$ are (anti-)parallel on $\lambda_{SPP}/2$ length scale (black scale bars). $Re(\mathbf{E}^* \cdot \mathbf{H})$ amplitude follows the envelope of the generation optical fields in time. **b,** The amplitudes and widths of $A_{ME}$ for the same spatial regions displayed in **a** for $\delta = 10^{-4}$ in Eq. 2. The arrows emphasize that the red and blue spots mark strong $A_{ME}$ for the clockwise and counterclockwise rotating SPP vortices in **a** with parallel and antiparallel $\mathbf{E}_{\parallel}$ and $\mathbf{H}_{\parallel}$ field orientations. The black solid curves indicate horizontal line profiles of $A_{ME}$ along the black dashed lines. The amplitudes and the full widths at half-maxima (FWHM) of $A_{ME}$ foci, ±12.6 and 10.3 nm, are the same for the circular and square CS. The thick solid lines mark the $\lambda_{SPP}/2$ (265 nm) scale.

The main result of our study is that vortex cores are foci of (anti-) parallel $\mathbf{E}_{\parallel}$ and $\mathbf{H}_{\parallel}$ fields, with orientations, which co-gyrate about the vortex core with the SPP frequency $\omega_L = 0.54$ PHz (vacuum $\lambda_L = 550$ nm; 2.25 eV photon energy; ~1.83 fs oscillation period) in the analytical calculations and experiments. The relative alignment of the co-gyrating $\mathbf{E}_{\parallel}$ and $\mathbf{H}_{\parallel}$ vortex fields is constant, and therefore can be maintained on Hz to PHz time scales over duration of the excitation pulse envelope. Femtosecond Movies S1 and S2 show the calculated co-gyration of the $\mathbf{E}_{\parallel}$ and $\mathbf{H}_{\parallel}$ fields in vortex cores generated by the circular and square symmetry SPP coupling structures[39] over one cycle of $\omega_L$; the images in Fig. 2a are defined by single frames extracted from such movies. Notably, the field gyration is defined by OAM $|l| = 1$ of the SPP field and is localized within a diameter of $\lambda_{SPP}/2$ centered on each vortex core, independent of the CS. This localization can be understood by the fact that the interference forming the vortex fields makes them stationary, such that the oscillation in space and time of the optical SPP excitation plane wave is converted to a local gyration of SPP fields in space at $\omega_L$. While plasmonic vortices have been studied in numerous publications, to the best of our knowledge this is the first description of this key, universal property, the parallel gyration of their **E** and **H** fields.

We further calculate the spatial distributions of $A_{ME}$ as shown in Fig. 2b. The magnetic polarizability being much smaller at optical frequencies than the electrical one, we assume $\delta = 10^{-4}$, because for $\delta \leq 10^{-4}$ the $A_{ME}$ distributions converge to the reported profiles[28]. For the single vortex and vortex array of the circular CS and square CS (Fig. 2b), the $A_{ME}$ profiles have the same amplitude and width suggesting that it is a universal property of $|l| = 1$ vortices. The circular CS has secondary circular sub-maxima of $\boldsymbol{E}^* \cdot \boldsymbol{H} \neq 0$ that are a consequence of the periodic **E**||**H** alignment seen in the relative angle plot in Fig. 1b.

The sign of vortex gyration and therefore $A_{ME}$ alternates between adjacent square domains for $Q = 4$. Despite the sign of $A_{ME}$ alternating from positive for clockwise to negative for counterclockwise gyration (indicated by the red and blue arrows and $A_{ME}$ densities in Fig. 2b), on a $\lambda_{SPP}/2$ length scale, the amplitude is constant over an optical cycle time scale. The spatial periodicity of $A_{ME}$ found for square vortex arrays, acquires additional structure for $Q > 4$, because of frustration of interaction of dipolar light with polygonal CS. Quasiperiodic SPP fields, e.g., for $Q \cong 5$, generate multiple vortex arrays of different vorticity[39]. Figure S1 shows exemplary $A_{ME}$ profiles for different sublattices of $Q = 6$ CS. What is notable, however, the central $A_{ME}$ peak profile is always the same independent of $Q$ (see Fig. S2). The black solid lines in Fig. 2b quantify the $A_{ME}$ spatial distributions by plotting profiles along the black dashed lines. The line profiles show that $A_{ME}$ foci are universal properties of plasmonic vortices with $|l| = 1$ that saturate for $\delta$ < $10^{-4}$ with FWHM of $\sim \frac{1}{51} \lambda_{SPP}$ (~10 nm), amplitude of ~12.6, and a sign that is negative (positive) for anti-(parallel) field alignments.

We further evaluate the chirality density $C_{ME}$ that is defined by $-2Im(\mathbf{E}^* \cdot \mathbf{H})$ of the SPP vortex fields[28]. This quantity is odd only with respect to the $\mathcal{P}$ symmetry and has a $\frac{\pi}{2}$ phase shift with respect to $A_{ME}$. As expected for evanescent waves[47], $C_{ME} \sim 0$ for SPP vortex fields (see Fig. S3), whereas in Fig. 2, the $\mathcal{PT}$ symmetric ME interaction dominates within vortex cores. I

### III. ULTRAFAST MICROSCOPY OF PLASMONIC FIELDS

We experimentally image the $\mathbf{E} \cdot \mathbf{H}$ fields and their $A_{ME}$ by performing ultrafast microscopy of plasmonic vortices when normally incident, circularly polarized light illuminates 2D silver films with lithographically inscribed polygonal SPP CS[36-38]. As described in Fig. 3, the lithographically formed grooves in CSs provide momentum to launch SPP fields with $k_{SPP}$ directed normal to their edges at silver/vacuum interface with the defined OAM. The circular polarization and CS geometry define SPP wavefronts that propagate at the speed of light towards their centers[39]. Depending on the CS geometry, SPP fields undergo photonic SOI[32] to form single, clusters, or arrays of plasmonic vortices with topological spin skyrmion[22,40] and meron[38,41,42] textures. The vortices form with a phase singularity at their core, where the phase of the surface normal $E_z$ field is undefined, and the $\mathbf{E}_\parallel$ and $\mathbf{H}_\parallel$ fields have maxima as they circulate in the *x-y* plane with a geometric phase winding of $2\pi$ [41,43,44] (Fig. A1), like water gyrating into a drain.

We generate movies of the SPP vortex field's nanofemto flow by interferometric time-resolved photoemission electron microscopy (ITR-PEEM). Two-photon photoemission signal (2PP) is excited by illuminating CSs with ~20 fs duration, $\lambda_L$=550 nm pulses generated by a non-collinear optical parametric amplifier (NOPA). The NOPA is pumped by the third harmonic of Yb-doped fiber oscillator/amplifier system (IMPULSE Clark-MXR) operating at a 1MHz repetition rate. The excitation pulses are sufficiently energetic to induce photoemission through nonlinear two photon excitation of electrons to above the work function of polycrystalline silver (~4.2 eV). An identical pulse pair with a variable delay $\tau$ defined with 100 attosecond precision is generated by passing the excitation light through a Mach-Zehnder interferometer. The pulse pairs illuminate the sample under ultrahigh vacuum to generate 2PP signal.

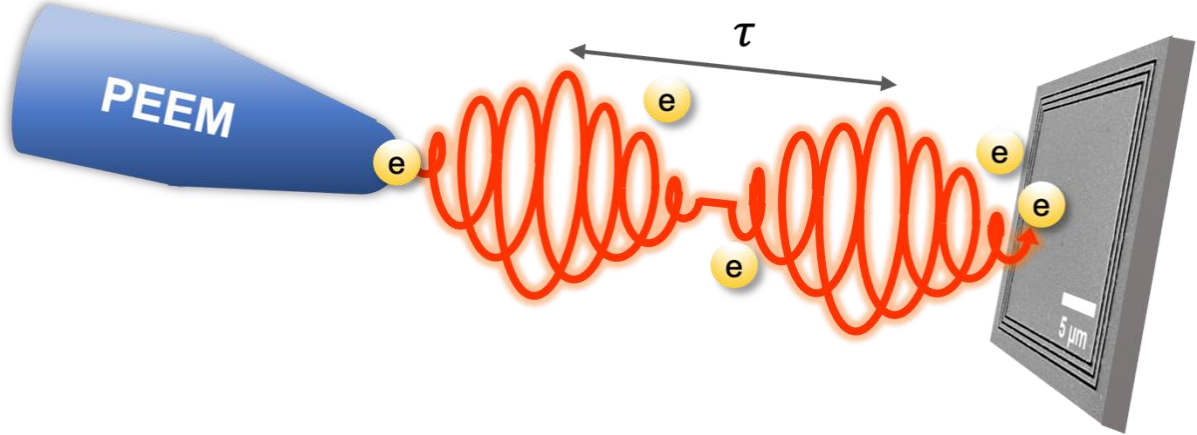

**Fig. 3: ITR-PEEM experiment set-up.** In a typical PEEM experiment, a coupling structure is etched on a metal surface to compensate for the momentum mismatch for launching SPP waves. The metal surface is excited by an externally incident laser pulse (optical field) to generate photoemission so that the photoemitted electrons can be detected. In an ITR-PEEM set up a pair of identical, phase-locked left circularly polarized fs pulses with pulse delay $\tau$ illuminates a silver film at a normal incidence. The silver film has lithographically inscribed square (or circular) structure for generation of SPP fields as shown in the scanning electron microscope image of a square slit coupling structure.

The CSs with square and circular geometry are nanofabricated in thermally evaporated ~100 nm thick Ag film on an n-type Si(111) substrate by focused ion beam lithography. The lithographic etching forms a slit of defined CS geometry having ~100 nm width and depth. The CS structure supplies the momentum to couple the optical field into emitting the SPP mode of Ag/vacuum interface with $k_{SPP}$ vector directed in the surface plane normal to the etched metal structure edge and having duration of the generating field[48]. Each CS consists of three concentric replicated slit trenches with the distance separating them increasing by $\lambda_{SPP} = 530$ nm, so that the generated SPP fields from each slit structure add in-phase to intensify the field (scanning electron microscopy image of a nanofabricated square CS shown in Fig. 3). In case of the circular (square) CS for generating the response in Fig. 4 (Fig. 5), as will be described next, each radius (side) of the inner-most coupling slit has the dimension of $5\lambda_{SPP}= 2.65$ μm ($30\lambda_{SPP}= 15.9$ μm).

In a PEEM experiment, single laser pulses excite 2PP signal from the sample which is proportional to $\int E_{total}^4(x,y,z,t)dt$ of the total electric field where $E_{total}(x,y,z,t) = E_{light}(x,y,z,t) + E_{SPP}(x,y,z,t)$, and $E_{light}$ and $E_{SPP}$ represent the incident optical field and the SPP field it creates, respectively[45]. In ITR-PEEM experiment, the pump-probe excitation creates additional $\tau$ dependent signal from the interference between the pump generated in-plane SPP wave packet and the optical field of the delayed probe pulse. The photoelectron spatial distributions imaged by the PEEM electron optics in $\Delta\tau = 100$ as steps generate nanofemto movies of the SPP field time evolution. The raw images captured by ITR-PEEM are processed by pixel-wise Fourier time filtering, to extract images that represent the $\tau$ dependent first-order optical field-SPP interaction consisting of interferences between their in-plane field components[38]. This measures in-plane SPP ***E*** field component parallel to the probe polarization, from which we calculate the other five ***E*** and ***H*** field components by Maxwell's equations[46] (Fig. S4). This enables one to extract vectorial electric and magnetic field flow of SPP wave packets with <50 nm spatial and 100 as time step resolution[45,46].

## IV. SUBDIFFRACTION LIMITED IMAGING OF MAGNETOELECTRIC DENSITY

Having established by simulation that the ME fields dominate within plasmonic vortex cores, for a single plasmonic vortex and an array of vortices of Fig. 2a and 2b, respectively, we confirm this expectation by imaging the SPP fields and extracting the nanofemto $A_{ME}$ distributions by ultrafast coherent microscopy technique described above.

### A. Ultrafast microscopy of a single plasmonic vortex

The single, central vortex formed by the circular CS has the same physical dimensions as calculated for all $Q$, and therefore can be taken as the fundamental property of plasmonic vortices. We first describe the associated ME response and then compare it to that of $Q = 4$ CS. As described by Dai *et al*.[22], the measured PEEM image of 2PP signal excited by a single pulse within the focus in a circular region of $\sim\lambda_{SPP}$ diameter centered at the SPP vortex core, is shown in Fig. S4a. In two pulse pump-probe excitation experiment on the same system, the individual $\boldsymbol{E}$ and $\boldsymbol{H}$ field components are extracted from movies of the 2PP distributions (see Fig. S4 and Supplemental

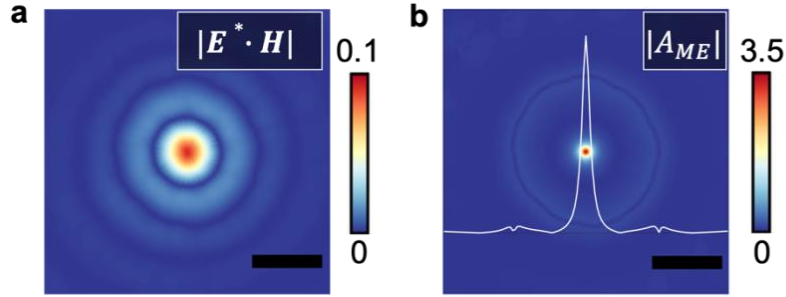

**Fig. 4: Ultrafast microscopy of *E·H* fields generated by a *Q* = ∞ CS. a,** The color scale represent $|\boldsymbol{E}^* \cdot \boldsymbol{H}|$ distribution from an ITR-PEEM data[22] surrounding an SPP vortex core generated by illuminating a polycrystalline Ag film with a circular CS with ultrafast LCP laser pulses (details in Supplemental Material). The measured $\boldsymbol{E}$ and $\boldsymbol{H}$ fields give a FWHM of the $|\boldsymbol{E}^* \cdot \boldsymbol{H}|$ distribution of $0.35\lambda_{SPP}$, and **b,** the ME density ($|A_{ME}|$) (amplitude represented by color scale). The white line in **b** is a horizontal line profile of $|A_{ME}|$ that quantifies the measured FWHM of 74 nm and height of 3.5 of $|A_{ME}|$. The line profile shows secondary maxima in $|A_{ME}|$ with a dip where the $\boldsymbol{E}$ and $\boldsymbol{H}$ field's orientations transition through orthogonal. The scale bars in **a-b** denote $\lambda_{SPP}/2$ (265 nm).

Material for further details). These field components define $|\boldsymbol{E}^* \cdot \boldsymbol{H}|$ and $|A_{ME}|$ distributions at the circular vortex core, as displayed in Fig. 4a and 4b, respectively. The measured primary $|\boldsymbol{E}^* \cdot \boldsymbol{H}|$ peak coincides with the vortex core having a Gaussian FWHM of $0.35\lambda_{SPP}$, and a skyrmion spin texture[22], while secondary maxima appear as surrounding rings. From Eq. 2, we further calculate the experimental $|A_{ME}|$ with a 74 nm FWHM and 3.5 height, which are broader and lower than the theoretically predicted values. This discrepancy can be attributed to the limited momentum range of low energy photoelectrons, which defines the PEEM resolution[49,50], and the sensitivity of PEEM imaging to environmental electric and magnetic fields. Significantly, the $|A_{ME}|$ distribution is an even function with respect to the origin, where the $E_z$ field vanishes at the vortex core, while the in-plane fields are maximum. Such field distributions are universal for the $|l| = 1$ single and multiple plasmonic vortex arrays that have the same profile in Fig. 2b.

**B. Ultrafast microscopy of a plasmonic vortex array**

Next, we analyze the experimental $E_\parallel$ and $H_\parallel$ fields of the plasmonic vortex array formed by square CS (Fig. 5). The experimental method is the same as for the data in Fig. 4, but the field component extraction is described in the Supplemental Material (see Fig. S5). The color scale in Fig. 5a shows evolution of the experimentally extracted $\boldsymbol{E}_\parallel \measuredangle \boldsymbol{H}_\parallel$ of the square vortex array by advancing $\tau$ in 1/8 of an-optical cycle steps (~0.23 fs). The detailed co-gyration of experimentally obtained $\boldsymbol{E}_\parallel$ and $\boldsymbol{H}_\parallel$ over 1/2 optical cycle is demonstrated in Fig. S6. Meanwhile, Fig. 5b reveals the experimentally obtained $Re(\boldsymbol{E}^* \cdot \boldsymbol{H})$ distribution, which is a constant within the experimental uncertainty. This experimentally verifies that at each vortex core, the $\boldsymbol{E}_\parallel$ and $\boldsymbol{H}_\parallel$ fields co-gyrate at the $\omega_L$~0.54 PHz optical frequency, while $Re(\boldsymbol{E}^* \cdot \boldsymbol{H})$ remains constant over an optical cycle.

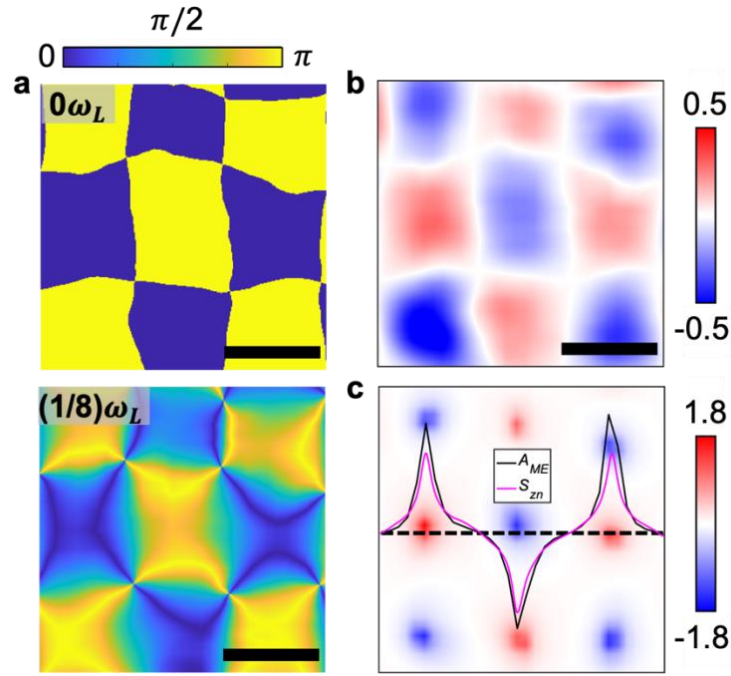

**Fig. 5: Ultrafast microscopy of *E*, *H*, and *E·H* fields for a square plasmonic vortex array. a-e,** Color maps represent the experimental relative $E_\parallel \measuredangle H_\parallel$ distribution evolving from 0 to 1/8 of an optical cycle (like Fig. 1). The vectorial $E_\parallel$ and $H_\parallel$ field components are extracted from the experimental data of a square array of SPP vortices[41] by procedure in Fig. S5 and shown in Fig. S6. **b,** The color map represents the experimental $Re(E^* \cdot H)$ distribution with red (positive) and blue (negative) values defined by respective parallel and antiparallel projections between $E_\parallel$ and $H_\parallel$. **c,** The extracted $A_{ME}$ distribution for the spatial region of **a-b** denoted through the red and blue color scheme. The sharp spots indicate strong localization of the ME driving fields at each plasmonic vortex core of the array shown in **a-b**. Note, that within an optical cycle the fields rotate at frequency $\omega_L$, the quantity $Re(\boldsymbol{E}^* \cdot \boldsymbol{H})$ remains unchanged within the experimental uncertainty. The black line in **c** shows the line profile of $A_{ME}$ taken along the black dashed line. The average heights and FWHMs of the peaks shown are 1.4 and 55 nm, respectively. For comparison, the purple line in the same plot denotes the line profile of the *z*-component of the normalized SAM ($S_{zn}$)[41] along the black dashed line. The scale bars in **a-b** indicate $\lambda_{SPP}/2$ (265 nm).

Thus, pulse duration and amplitude modulate the strength of the ME density, while $\omega_L$ defines the driving frequency of spin and charge excitations. Fig. 5c shows the distribution of $A_{ME}$ from the experimental data assuming $\delta = 10^{-4}$ for the same spatial region of Fig. 5a-b. The experimental (Fig. 5c) and calculated (Fig. 2b) $A_{ME}$ profiles agree qualitatively, though the limited momentum range of photoelectrons, and fluctuations in experimental conditions broaden its predicted focusing, which is at the limit of the specified PEEM resolution[49,50]. The horizontal line profile of $A_{ME}$ (continuous black line) along the black dashed line in Fig. 5c, shows $A_{ME}$ to be sharply peaked at vortex cores with alternating signs, having average amplitudes and FWHMs of 1.4 and ~55 nm, respectively, that verify the deep sub-diffraction localization of the ME driving fields at plasmonic vortex cores. The line profile of the out-of-plane (*z*) component of the normalized SAM ($S_{zn}$)[41] along the black dashed line is indicated in the same plot through the purple line. This reveals the alignment of $A_{ME}$ and $S_{zn}$ at the vortex core, which is expected because the in-plane SPP fields circulate around the cores.

## V. CONCLUSIONS

The vectorial nature of SPP fields, and photoelectron-based microscopy enables to create and image plasmonic $E \cdot H$ field textures in a phase-resolved manner far beyond the Abbe diffraction limit of conventional optical microscopy. Our examination of structured vortex fields invites further examination of quantum vortex fields in superconductors, superfluids, topological states of matter, Bose-Einstein condensates, vacuum structured light, [51-55] and their interactions with topological electromagnetic fields[56]. While the ME interaction has been detected by application of DC magnetic fields through the Faraday and Kerr polarization rotations[25], the finite divergence of the $H$ field in structured light enables the application of $E \cdot H$ fields at THz-PHz frequencies to examine coherent magnetless nonreciprocal phenomena in the quantum regime[57]. Beyond our discovery and space-time imaging of $E \cdot H$ focus based on ultrafast microscopy of plasmonic vortices, we anticipate that other schemes may similarly structure $E$ and $H$ fields, as has already been demonstrated for plasmonic nanoparticles[21], and similar effects are likely to be observed in propagating optical fields with topological field textures[56]. Therefore, the application of structured light in a laboratory setting may have broad applications for study of linear and nonlinear magnetoelectric spin-charge coherent photonics[58] such as dynamical symmetry breaking in topological physics, non-reciprocal and non-Hermitian physics[59,60], and exploration of the origin of homochirality[11]. At plasmonic vortex cores, we estimate that for the circular CS we focus parallel ~1 GV/m electric and ~10 T magnetic fields within a gyrating core of $0.35\lambda_{SPP}$ FWHM diameter. The plasmonic field is likely to be further enhanced by nonlocal screening response and confinement to ~330 nm above a silver plasmon supporting surface, where it can evanescently dress proximate quantum materials[23]. The ME density of vortex fields establishes them as highly localized source-drain of both fundamental cosmological axions and synthetic axionic quasiparticles in strong topological materials. While experiments aiming to detect cosmological axions rely on large volumes to detect the weak axion interactions[61], the ability to focus $A_{ME}$ on ~10 nm length and 20 fs (pulse duration limited) time scales with the circular CS provides for deep sub-diffraction microscopy of ME phenomena and $\mathcal{PT}$ symmetry interactions[8,62,63]. In the case of the square vortex array, the sign alternation of $A_{ME}$ of the oppositely gyrating vortices in adjacent cells can drive the spin and charge currents to generate spin polarized edge flows for finite sized CS. We anticipate that near-PHz frequency gyrating $\boldsymbol{E \cdot H}$ fields of plasmonic vortices can drive in their nearfield the nonconservative fluctuations of the coupled spin-charge collective plasmonic responses in Weyl semimetals[15,64], enabling the study of chiral anomaly and other topological effects in condensed matter. The space-time focus of ME interactions is on the Fermi surface electron scattering length and time scales in normal metals, and thus the coherent femtosecond ITR-PEEM microscopy is well positioned to image dissipationless currents, for example directional dichroism, associated with axion physics[10]. The topological protection, and the ability to excite and probe ME densities offers new ways to interact with quantum matter opening the window for exploring axion physics[65] in a laboratory setting.

**ACKNOWLEDGEMENTS**

We acknowledge the valuable research resources from the University of Pittsburgh Center for Research Computing, and the Nanoscale Fabrication and Characterization Facility. W.V.L thanks AFOSR Grant No. FA9550-23-1-0598, the MURI-ARO Grant No. W911NF17-1-0323 through UC Santa Barbara, and the Shanghai Municipal Science and Technology Major Project through the Shanghai Research Center for Quantum Sciences (Grant No. 2019SHZDZX01). Y.D. thanks the National Natural Science Foundation of China (Grant No. 12374223) and Shenzhen Science and Technology Program (Grant No. 20231117151322001). H.P. thanks the R. K. Mellon foundation for support. A.G. thanks the Quantum Science and Engineering Fellowship from the Pittsburgh Quantum Institute. All the authors thank A. Bhoonah and B. Batell for participating in discussions of axion physics.

**APPENDIX A: DETAILS OF ANALYTICAL MODEL**

The simulations of the SPP $\boldsymbol{E}$ and $\boldsymbol{H}$ field distributions in Figs. 1 and 2 is performed by an analytical model for interfacial electromagnetic fields [22,38,40,41]. When a lithographically defined continuous, homotopic polygonal coupling structure (CS) engraved in a polycrystalline Ag film is illuminated with normally incident, circularly polarized optical pulses, it generates SPP fields that are steered into a linear superposition to form a vortex interference pattern. Constructive interferences of topological SPP fields that give strong 2PP signal are defined by the CS geometry and the optical polarization of the excitation light. Sharp geometrical edges cut into silver films to form CSs provide momentum that couples the incident optical fields into the SPP mode of the Ag/vacuum interface with $k_{SPP}$ aligned normal to the CS structure edges. The optical polarization and CS geometry define the phase and the OAM of the SPP field. In the analytical model we consider left circularly polarized (LCP) light with spin angular momentum (SAM), $\sigma = -1$, interacting with a continuous CS with geometric charge, $m = 0$, to create SPP wave packet with OAM, $l = m + \sigma$ propagating in the *x-y* plane. Right circularly polarized light would generate the opposite vortex circulation on account of $\sigma = +1$. The OAM defines the topological charge of the generated SPP wave packets that undergo SOI of light. Specifically, fields are expressed as a superposition of SPP waves in the *x-y* plane launched from different sides of a polygon with $Q$ edges under the excitation of with light helicity $\sigma = -1$ such that the transverse (*z*) component of SPP electric field is[39],

$$E_z(x,y,z,t) = E_0 \sum_{j=1}^{Q} e^{i\varphi_j} e^{-ik_{SPP}[\sin(\psi_j)x+\cos(\psi_j)y]} e^{-Kz} e^{-i\omega t}, \quad (A1)$$

where $E_0$ is a constant, $\psi_j = -\frac{2\pi j}{Q}$ is the relative angle of $k_{SPP} = \frac{2\pi}{\lambda_{SPP}}$ emerging from each polygon side with respect to the *y* axis, and $\varphi_j = \frac{2\pi j\sigma}{Q}$ is the relative phase of the $j^{th}$ SPP wave and $K$ is the decay constant in +*z* direction. It is essential to note that equation (A1) denotes the $E_z$ field normal to the *x,y* plane. The SPP pattern forms through interference of *j* components with appropriate $k_{SPP}$ vectors, polarizations, and phase differences. At the Ag/vacuum interface ($z = 0$), the *x,y* dependence of the $E_z$ the field of equation (A1) can be rewritten as,

$$E_z(x,y) = E_0 \sum_{j=1}^{Q} e^{i\varphi_j} e^{-ik_{SPP}[\sin(\psi_j)x+\cos(\psi_j)y]}. \quad (A2)$$

By Maxwell's equations, the remaining field components are given by,

$$E_x(x,y) = E_0 \sum_{j=1}^{Q} i\frac{K}{k_{SPP}} \sin(\psi_j) e^{i\varphi_j} e^{-ik_{SPP}[\sin(\psi_j)x+\cos(\psi_j)y]}, \quad (A3)$$

$$E_y(x,y) = E_0 \sum_{j=1}^{Q} i\frac{K}{k_{SPP}} \cos(\psi_j) e^{i\varphi_j} e^{-ik_{SPP}[\sin(\psi_j)x+\cos(\psi_j)y]}, \quad (A4)$$

$$H_x(x,y) = E_0 \sum_{j=1}^{Q} \frac{Kk_{SPP}}{\omega} \cos(\psi_j) e^{i\varphi_j} e^{-ik_{SPP}[\sin(\psi_j)x+\cos(\psi_j)y]}, \quad (A5)$$

$$H_y(x,y) = E_0 \sum_{j=1}^{Q} -\frac{Kk_{SPP}}{\omega} \sin(\psi_j) e^{i\varphi_j} e^{-ik_{SPP}[\sin(\psi_j)x+\cos(\psi_j)y]}. \quad (A6)$$

For transverse magnetic waves, $H_z = 0$. Equations (A3) – (A6) show that $\boldsymbol{H}_\parallel = (H_x, H_y)$ and $\boldsymbol{E}_\parallel = (E_x, E_y)$ have the same *x,y* dependence, though their sign may be the same or the opposite depending on the local field phase.

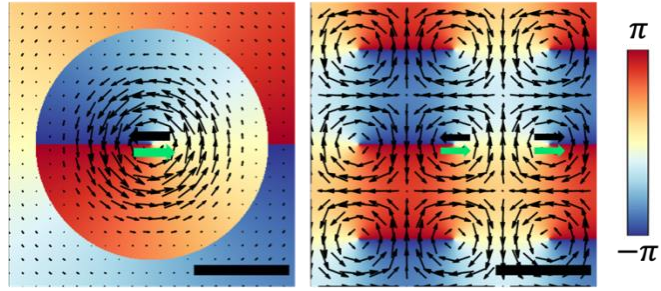

**Fig. A1: The SPP vortex fields.** Color maps showing the phase angle variation from $-\pi$ to $\pi$ radians of the $E_z$ component of SPP field when generated by illuminating circular (left) and square (right) CS with LCP light for the same spatial region that is shown in Fig. 2. The black arrows show the wavevector $k_{SPP}$ of the field circulation. At the center of each circulation there is a phase singularity where $E_z = 0$ and the phase is undefined. The thick black and green arrows indicate the instantaneous directions of the rotating $\boldsymbol{E}_\parallel$ and $\boldsymbol{H}_\parallel$ SPP fields at the vortex cores, respectively, as illustrated in Fig. 2a. The scale bars indicate $\lambda_{SPP}/2$ (265 nm). The details of $m = 0$ plasmonic vortex fields can be found in Refs. [22,39,41].

Fig. A1 shows a calculated map of the $E_z$ field phases for illumination of the circular and square CSs, based on the above-described analytical model. The black arrows show $k_{SPP}$ of circulating SPP vortex energy flow. For the circular CS there is a single vortex, while for the square CS there is an array of vortex fields with opposite circulation in alternating $\lambda_{SPP}/2$ square domains. For the array that is generated by the $Q = 4$ structure, at each vortex core, the SPP $\boldsymbol{E}_\parallel$ (thick black arrow) and $\boldsymbol{H}_\parallel$ (thick green arrow) fields are parallel (clockwise vortex circulation) or antiparallel (counterclockwise vortex circulation), as discussed in Fig. 2.